\documentclass[aps,prl,twocolumn,showpacs,superscriptaddress]{revtex4}
\usepackage{epsfig}
\def\avg#1{\langle#1\rangle}

\def\be{\begin{equation}}       \def\ee{\end{equation}}
\def\bea{\begin{eqnarray}}      \def\eea{\end{eqnarray}}

\def\avg#1{\langle#1\rangle}

\def\be{\begin{equation}}       \def\ee{\end{equation}}
\def\bea{\begin{eqnarray}}      \def\eea{\end{eqnarray}}

\begin{document}
\def \cH{{\cal H }}
\def\nd{{^{\vphantom{\dagger}}}}
\def\yd{^\dagger}
\def \bea{\begin{eqnarray}}
\def \eea{\end{eqnarray}}

\title{The Helical Liquid and the Edge of Quantum Spin Hall Systems}
\author{Congjun Wu}
\affiliation{Department of Physics, Stanford University, Stanford,
CA 94305}
\affiliation{Kavli Institute for Theoretical Physics, University
of California, Santa  Barbara, CA 93106}
\author{B. Andrei Bernevig}
\affiliation{Department of Physics, Stanford University, Stanford,
CA 94305}
\author{Shou-Cheng Zhang}
\affiliation{Department of Physics, Stanford University, Stanford,
CA 94305}

\begin{abstract}
The edge states of the recently proposed quantum spin Hall systems
constitute a new symmetry class of one-dimensional liquids dubbed
the ``helical liquid'', where the spin orientation is determined 
by the direction of electron motion. We prove a no-go theorem 
which states that a helical liquid with an odd number of components cannot be
constructed in a purely $1$D lattice system. In a helical liquid
with an odd number of components, a uniform gap in the ground state
can appear when the time-reversal (TR) symmetry is spontaneously
broken by interactions. On the other hand, a correlated two-particle
backscattering term by an impurity can become relevant
while keeping the TR invariance. The Kondo effect in such
a liquid exhibits new features in the structure of the
screening cloud. 
\end{abstract}
\maketitle

The field of spintronics is largely 
motivated by the possibility of
low power logic devices designed using the spin degree of
freedom of the electron \cite{wolf2001}. 
Recently, it has been proposed that in semiconductor materials with
spin-orbit (SO) coupling a dissipationless spin current can be induced  by an
electric field \cite{murakami2003}.
The theoretical prediction of this ``intrinsic spin Hall effect
(SHE)"\cite{murakami2003,sinova2004} has stimulated tremendous
research activity both theoretical and experimental.
On the theoretical side, it has been shown that the vertex correction due to
impurity scattering vanishes in the $p$-doped Luttinger 
and Rashba models\cite{murakami2004,bernevig2005c}, 
while it actually cancels
the intrinsic SHE in the $n$-doped Rashba
model\cite{inoue2004}. 
Recent experimental results in the GaAs system with both electron
and hole doping \cite{kato2004,wunderlich2005} are consistent with the
existence of SHE although more work is necessary to
determine the intrinsic versus extrinsic nature of the observed effect.

However, the electric field in the SHE systems still generates the 
ohmic dissipation in the charge channel  of a doped semiconductor.
This issue motivated the proposal of a spin Hall
insulator \cite{murakami2004a}, where the spin current is not
accompanied by the charge current. More recently, the quantum SHE
(QSHE) has been proposed in systems with \cite{bernevig2005} or
without Landau levels \cite{haldane1988,kane2004,qi2005,onoda2005}.
The QSHE has as a central concept the existence of a bulk gap 
and gapless edge states in a TR invariant system with SO coupling. 
In the ideal QSHE, the  left-movers on the edge are correlated
with down-spin $\downarrow$, the right-movers have
up-spin $\uparrow$, and the transport is quantized. 
We dub these edge states  a ``helical liquid", 
which describes the correlation between the spin and the momentum. 
As spin is not conserved, the extra SO interactions (e.g. Rashba)
change the quantized nature of the ideal system.
However, the edge transport turns out
to be quite robust: as long as the bulk gap is not closed, numerical
results find that the spin Hall conductance remains near the
quantized value, being rather insensitive to disorder scattering,
until the energy gap collapses with increasing SO
coupling \cite{sheng2005}.

The helical liquid constitutes a new symmetry class of 1D
liquids; unlike the chiral Luttinger liquid, it does not break TR
invariance and unlike the usual spinless Luttinger liquid, whose TR
transformation satisfies $T^2=1$, the TR transformation of the
helical liquid satisfies $T^2=-1$. 
Unlike the spinful Luttinger liquid, 
the spinful Luttinger liquid has to have an even number of components
(branches of TR pairs),
while the helical liquid can have an odd number of components.
Recently, Kane and Mele\cite{kane2005} pointed out that helical liquids 
with an even or an odd number of TR components are topologically distinct 
and are characterized by a $Z_2$ symmetry in the non-interacting case.
However, this work shows that in the presence of strong interactions,
no strict topological distinctions between the even and odd helical liquids
exist, while important quantitative differences do remain.

In this letter, we analyze the properties of the helical liquid. We
write down the lattice Hamiltonian and 
show that the fermion doubling theorem proves 
that the helical liquid with an odd number of
components cannot be constructed in purely 1D lattices and hence
must arise as an edge effect of a bulk 2D lattice.
In that sense, such a 1D
helical liquid must be a ``holographic liquid". We then analyze the
effects of TR invariant interactions.
The Umklapp term can potentially open up a gap
by spontaneously breaking the TR symmetry 
in the ground state.
Disordered two-particle backscattering also induces a glass-like TR
breaking ground state. Fortunately, both  cases require extreme
repulsive interactions which are unlikely to be experimentally
realized. 

We begin with the QSHE in the Landau level picture \cite{bernevig2005}, which
can be understood as two opposing effective orbital magnetic
fields $\pm B$, realized through a special position-dependent SO coupling,
acting on spin $\uparrow$ and $\downarrow$ electrons.
On a lattice $(x,y) = (m,n)$, with an edge on the $x$-axis, and in
the Landau gauge simulated by a linear strain gradient in a sample
grown on the $[110]$ direction \cite{bernevig2005}, the
Schr\"{o}dinger equation now becomes:
\begin{eqnarray}
&  E \psi_m(k_y)= -t_x (\psi_{m+1}(k_y) + \psi_{m-1}(k_y)) - \nonumber \\
& \left(%
\begin{array}{cc}
 2 t_y \cos(k_y - 2\pi m \phi) & 2 \alpha \sin(k_y) \\
  2 \alpha \sin(k_y) & 2 t_y \cos(k_y + 2 \pi m \phi) \\
\end{array}%
\right) \psi_m (k_y) \nonumber
\end{eqnarray}
\noindent with $\psi_m = (\psi_{m, \uparrow}, \psi_{m,
\downarrow})^T$ a $2$-component spinor,  flux per
plaquette $\phi = p/q$, with $p,q$ relatively prime integers. An
extra Rashba SO coupling interaction $-2 \alpha k_y
\sigma_x$, which mixes the two spins has been added. By using the
transfer matrix formalism \cite{hatsugai1993} we can however prove
that for $\alpha \ll t_x, t_y$ , 
and to order $\alpha$, the edge states
do not differ in eigenvalues from the $\alpha =0$ states. There are
$q -1$ edge states and each of them is doubly degenerate. The
degeneracy is removed to order $\alpha^2$. 

\begin{figure}
\centering\epsfig{file=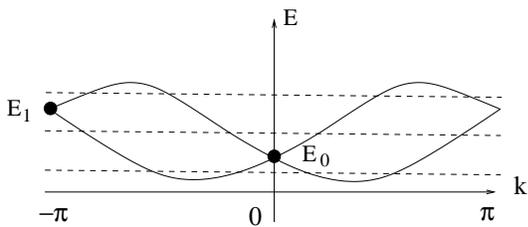,clip=1,width=0.8\linewidth}
\caption{The band structure in 1D lattice systems with time-reversal
symmetry. Even numbers of Kramers doublets appear at a
given energy
except at the extremum points. }\label{fig:oddpair}
\end{figure}

A generalization of the no-go theorem for 
chiral fermions in lattice models \cite{Nielsen1981}
can be used to prove that it is impossible to construct a purely
1D lattice model of the helical liquid with an odd number of
components \cite{kane2004,bernevig2005}.
We first consider the one-component
model with TR symmetry, i.e., two states with orthogonal spin
configurations for each momentum $k$ in the Brillouin zone (BZ).
The Kramers theorem associated with  $T^2=-1$  ensures
that eigenstates with $k=0$ or $\pi$ are doubly degenerate.
Their energies are denoted by $E_{0}$ and $E_1$, respectively.
Without loss of generality, we assume $E_1>E_0$.
TR symmetry also ensures $E_\sigma(k)=E_{\bar \sigma}(-k)$ for
Kramers doublets with opposite momenta and orthogonal spins.
As a result, two dispersion curves start from $k=-\pi$ and converge at $k=0$
in the left half of  the BZ .
In the right half of the BZ, the curves are symmetric to those in
the left half by a mirror reflection.
If an  energy $ E$ satisfies $E_0<E<E_1$, it crosses each
branch in the left half of the BZ an odd number of times.
On the other hand, an energy with $E>E_1$ or $E<E_0$ crosses
each branch in the left half of the BZ an even number of times. 
Thus for any general energy $E$, unless it is a local
energy maximum or minimum, the total number of crossing points is even in the
left half of the  BZ as shown in Fig. \ref{fig:oddpair}.
Thus the purely 1D band structure generally gives the helical liquid with
an even number of components.
This result can be straightforwardly generalized to the multi-band case.
However, the helical liquid with an odd number of components
can appear in the edge of
a 2D system, because the edge states do not necessarily cover the
entire BZ.

TR invariance imposes a strong constraint
on the correlation function between Kramers pairs.
Regardless of interactions and disorder, the
Green's function is: $G(\vec rt\downarrow; \vec r t^\prime\uparrow)=
\avg{\psi_\uparrow(\vec r t) \psi^\dagger_\downarrow(\vec r^\prime t^\prime
)}=0$, where $\avg{}$ means thermal average \cite{meir1993}.
To generalize this result to multiple pairs,
we define a set of fermion annihilation operators of
Kramers pairs as $\hat \psi_i$ and $\hat{\bar
\psi}_i~(i=1,2,..n)$ which satisfy $ T^{-1} \hat \psi_i T=
\hat{\bar \psi}_i$ and  $T^{-1} \hat{\bar \psi}_i T= -\hat \psi_i$.
Their  $2n$-point correlation functions are defined as
\bea
G_n(t_1,t_2,...t_n;t_n^\prime,...,t_2^\prime,t_1^\prime)&=&
\avg{ \hat \psi_1(t_1) \hat \psi_2(t_2)... \hat \psi_n(t_n)
\nonumber \\
&\times&
\hat {\bar \psi}^\dagger_n(t^\prime_n) ...
\hat {\bar\psi}^\dagger_2(t^\prime_2) \hat {\bar
\psi}^\dagger_1(t^\prime_1) }. \ \ \  \ \ \
\eea
$T^2=-1$ ensures that
$G_n(t_1, t_2, ...t_n; t_n^\prime,...t^\prime_2,t_1^\prime)
= (-)^n G(-t^\prime_1, -t^\prime_2...
-t^\prime_n;-t_n,...-t_2,-t_1)$.
Combining this with the time translational symmetry, we obtain
\bea
G_n(t,t,..t;0,0,..0)=0 \ \ \ (\mbox{n is odd})\label{eq:1TRpair},
\eea
regardless of interactions and disorder.
With the interpretation of $\psi$ and $\bar\psi$ as the right and left
movers,
the observation of Kane and Mele \cite{kane2005} that the single particle
backscattering is forbidden is still correct even in the presence of
interactions. However, this does not necessarily mean that the
system is gapless. In fact, a two-particle correlated backscattering
is  allowed  as
\bea G_4^\prime=G_4(t,t; 0,0) +\avg{\hat
\psi_1(t) \hat{\bar \psi}^\dagger_2(0)} \avg{ \hat \psi_2(t)
\hat{\bar \psi}^\dagger_1(0)}, \label{eq:2TRpair} 
\eea
which
effectively describes the propagation of a composite boson, and can
have nonzero values.

Next we discuss the interaction effects in the one-component model,
paying special attention to
the two-particle correlated backscattering in Eq.
\ref{eq:2TRpair}. For simplicity, we consider the band structure
with conserved $s_z$ \cite{bernevig2005,kane2004,qi2005}.
We linearize the spectra and arrive at
the non-interacting part as
\bea
H_0&=& v_f \int d x  ( \psi^\dagger_{R\uparrow} i \partial_x \psi_{R\uparrow}
-\psi^\dagger_{L\downarrow} i \partial_x \psi_{L\downarrow} ),
\label{eq:ham0} \eea
where the right (left) movers
$\psi_{R\uparrow} (\psi_{L\downarrow})$ carry spin up (down)
respectively.
The single particle backscattering term is just the $2k_f$ spin
density wave operators
$ N_x=\psi^\dagger_{R\uparrow} \psi_{L\downarrow}+ h.c.$
or $ N_y=i(\psi^\dagger_{R\uparrow} \psi_{L\downarrow}- h.c. )$,
which open up a mass gap in the spinless Luttinger liquid, is not
allowed here by virtue of being TR odd.
In contrast, for an even number of Kramers pairs
$\psi_{iR \uparrow}, \psi_{iL\downarrow} (i=1\sim n)$,
it is easy to write down a TR invariant mass term.
For $n=2$, a possible term is $\psi^\dagger_{1R \uparrow}
\psi_{2L \downarrow} - \psi^\dagger_{1R \downarrow} \psi_{2R
\uparrow} + h.c.$
Eq. \ref{eq:ham0} is also different from the Luttinger liquid
with the Rashba SO coupling \cite{gritsev2005} where
two branches of Kramers pairs still exists.

Only two TR invariant non-chiral interactions are allowed: the forward
and Umklapp scatterings 
\bea H_{fw}&=& g \int dx
\psi^\dagger_{R\uparrow} \psi_{R\uparrow}
\psi^\dagger_{L\downarrow} \psi_{L\downarrow} \\
\label{eq:fw} H_{um}&=& g_u \int dx  e^{-i 4 k_f x}
\psi^\dagger_{R\uparrow} (x) \psi^\dagger_{R\uparrow} (x+a)
\psi_{L\downarrow} (x+a) \nonumber \\
&\times& \psi_{L\downarrow}(x) +h.c. , \label{eq:umklapp} \eea
where a point splitting with the lattice constant $a$.
The chiral interaction terms only renormalize
the Fermi velocity, and thus are ignored. The $H_{um}$ term flips
two spins simultaneously, which can be microscopically obtained
from anisotropic spin interactions such as $\sum_{\avg{ij}} s_x(i)
s_x(j)-s_y(i) s_y(j)$ or $\sum_{\avg{ij}} s_x(i) s_y(j)+s_y(i)
s_x(j)$. 
The terms of $H_0$ and $H_{fw}$ 
have an $SO(2)$ rotational symmetry around the
$z$-axis, which is further explicitly 
broken by $H_{um}$ to $Z_2$ , i.e. $s_{x,y}\rightarrow -s_{x,y},
 s_z\rightarrow s_z$.

\begin{figure}
\centering\epsfig{file=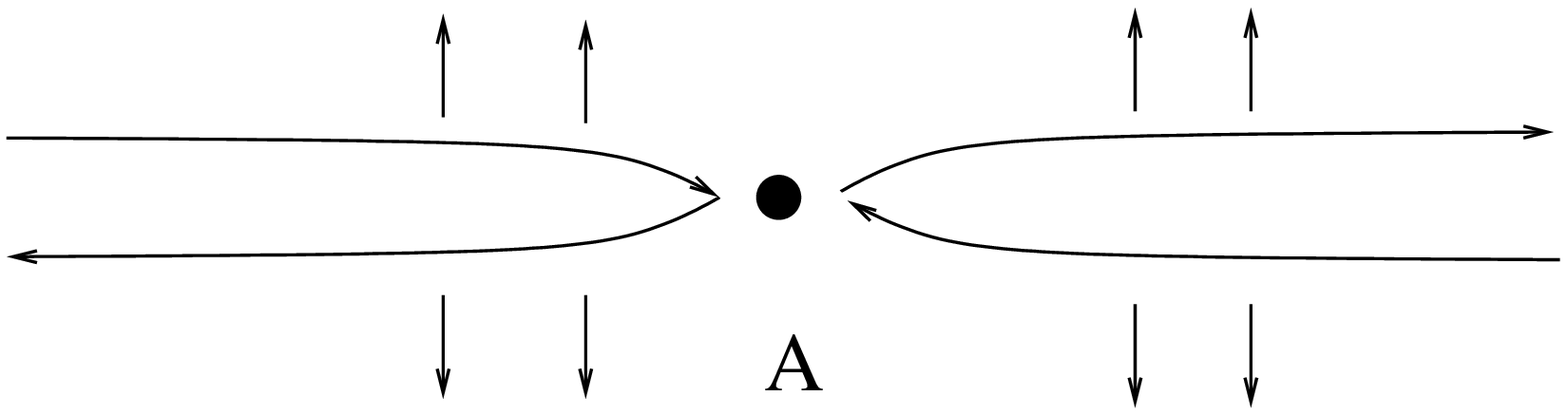,clip=1,width=0.8\linewidth}
\centering\epsfig{file=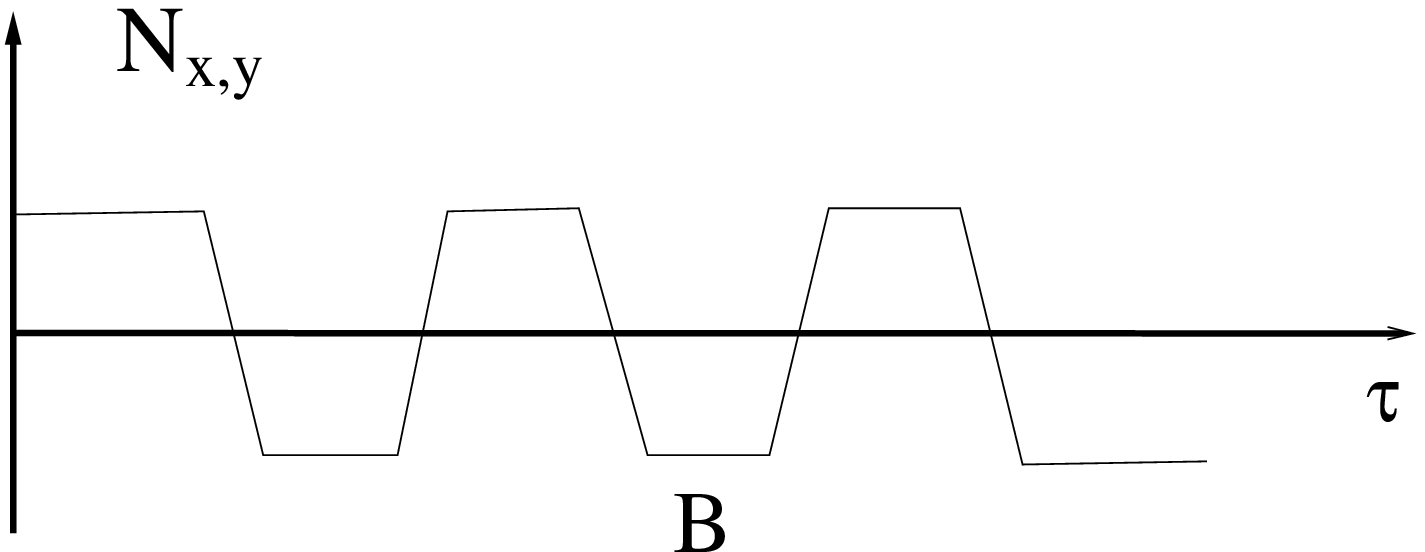,clip=1,width=0.7\linewidth}
\caption{A): The two-particle correlated backscattering process at
$K<1/4$ divides the line into two half-lines. B) The instanton
process of  $N_{x,y}(0,\tau)$
restores the TR symmetry. }\label{fig:back}
\end{figure}

It is well known that the forward scattering term gives the nontrivial
Luttinger parameter through $K=\sqrt{(v_f-g)/(v_f+g)}$, but still
keeps the system gapless. Only the Umklapp term has the potential to
open up the gap at commensurate filling $k_f=\pi/2$.
The standard bosonized Hamiltonian
reads
\bea
H&=& \int d x \frac{v}{2}\Big\{ \frac{1}{K} (\partial_x
\phi)^2 + K (\partial_x \theta)^2 \Big\}
+ \frac{g_u \cos \sqrt {16 \pi} \phi }{2 (\pi a)^2}, \nonumber \\
\eea
where $v=\sqrt{v_f^2-g^2}$ is the renormalized velocity,
$\phi=\phi_R+\phi_L$ and $\theta=\phi_R-\phi_L$ respectively.
$\phi$ contains both spin and charge degrees of freedom, and is
equivalent to the combination of $\phi_c-\theta_s$ in the spinful
Luttinger liquid. It is also a  compact variable
with a period of $\sqrt \pi$. The standard renormalization
group (RG) analysis shows that the Umklapp term becomes relevant
at $K_c<1/2$ with a pinned value of
$\phi$.
Consequently, a gap $\Delta\approx a^{-1} (g_u)^{\frac{1}{2-4K}}$
opens and the spin transport is blocked.
The mass order parameters $N_{x,y}$
whose bosonized forms are $ N_x=
\frac{i\eta_R \eta_L} {2\pi a} \sin \sqrt{4\pi} \phi, N_y=
\frac{i\eta_R \eta_L} {2\pi a} \cos \sqrt{4\pi} \phi,$ 
are odd under TR transformation.
At $g_u<0$, $\phi$ is pinned at either $0$ or $\sqrt\pi/2$, thus
the $N_y$ order is Ising-like.
At $T=0K$, the system is in the Ising-ordered phase, thus
TR symmetry is spontaneously broken.
On the other hand, when $0<T\ll \Delta$, $N_y$ is disordered
in 1D, thus the gap remains and TR
symmetry is restored by thermal fluctuations.
Similar reasoning applies to the case of $g_u>0$
where $N_x$ is the order parameter.

Now we consider the case that the Umklapp term
only exists in one single bond.
It then behaves as  an impurity-induced 2-particle
correlated backscattering term \bea H_{bs}^\prime&=& \int dx \ \ \
\delta(x) \frac{g_u }{2 (\pi a)^2}\cos \sqrt {16 \pi} \phi (x),
\label{eq:impurity}
\eea
as depicted in Fig. \ref{fig:back} A.
This boundary Sine-Gordon (BSG) term was studied in Ref.
\cite{Kane1992}. 
It can be reduced to a single-particle problem with the
effective action for $\phi(x=0,\tau)$  equivalent to a 1D
classical Coulomb plasma problem \cite{Kane1992}.
The RG analysis shows that the BSG term becomes relevant at $K<1/4$.
In this case, the 1D line is broken into two separated half-lines,
thus it is insulating for charge transport along the line.
However, it remains gapless and can support spin transport.
Because the BSG term only exists in a small region,
TR symmetry cannot be spontaneously broken.
Without loss of generality, we assume
$g_u\rightarrow - \infty$ in the RG process, then
Eq. \ref{eq:impurity} has two energy minima $\phi(0,\tau)
=0,\sqrt\pi/2$, which give $N_y(\tau)$ the same finite
amplitude but with opposite signs.
As a result, an electron can be back-scattered by
flipping its spin.
The instanton events in Fig. \ref{fig:back} B ,
i.e., the tunneling processes between these two classical minima,
restore TR symmetry.
Similar reasoning also applies to the case of $g_u\rightarrow +\infty$.

Now we discuss the two-particle backscattering due to quenched
disorder, described by the term:
\bea
H_{dis}&=& \int dx  \
\ \ \frac{ g_u(x) }{2 (\pi a)^2}\cos \sqrt {16 \pi} (\phi
(x,\tau)+\alpha(x) ) \ \ \,
 \eea
where the 
scattering strength $g_u(x)$ and phase $\alpha(x)$ are Gaussian
random variables. The standard replica analysis shows the disorder
becomes relevant at $K<3/8$ \cite{GIAMARCHI1988,XU2005}. Then in the
ground state  $N_{x,y}(x)$ show a glassy behavior
which is disordered in the spatial direction but static in the
time direction. Thus the spin transport is blocked and
TR symmetry is again spontaneously broken at $T=0K$.
Again at very low but finite $T$, the system remains gapped
with TR symmetry restored.

In the above, we have shown that
the helical liquid could {\it in principle} open up
a gap without breaking the TR symmetry at very low temperatures,
we conclude that there is no strict topological
distinction \cite{kane2005} between the helical liquid with an even
or odd number of components in the presence of disorder and
interactions.
However, for a reasonably weak interacting system with $K\approx 1$,
the one component helical liquid  remains gapless.

\begin{figure}
\centering\epsfig{file=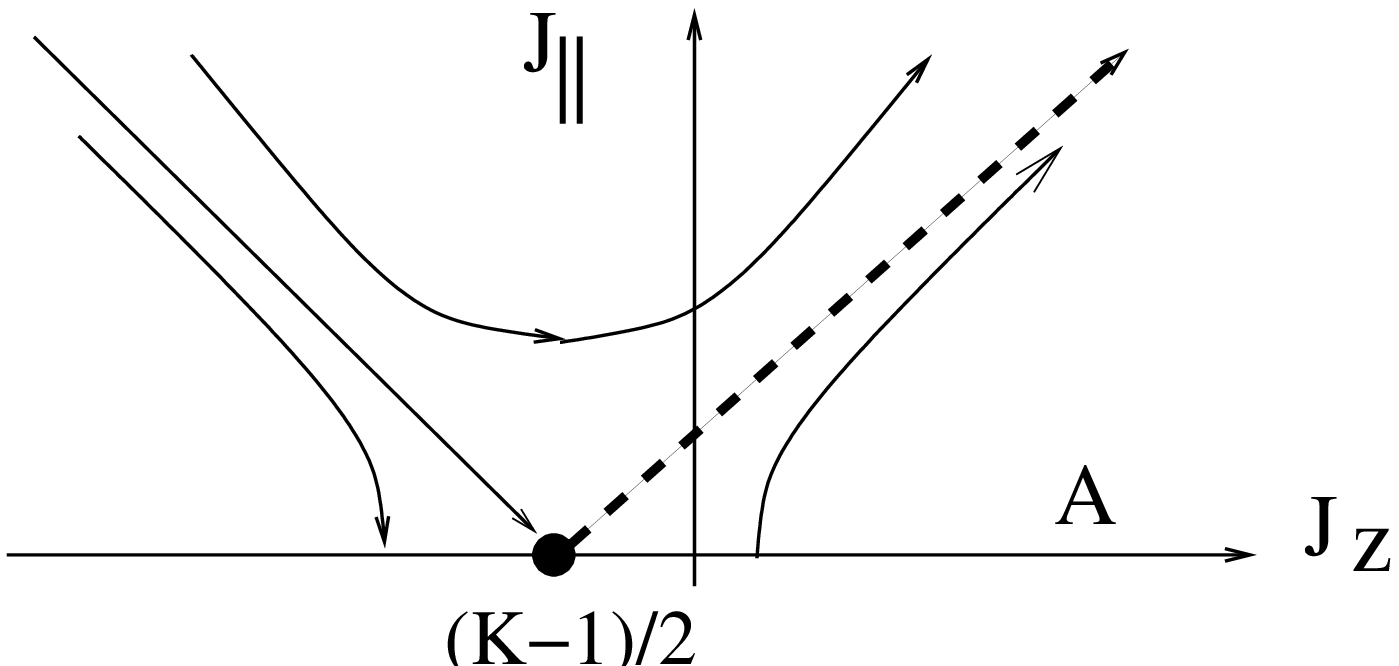,clip=1,width=0.7\linewidth}
\centering\epsfig{file=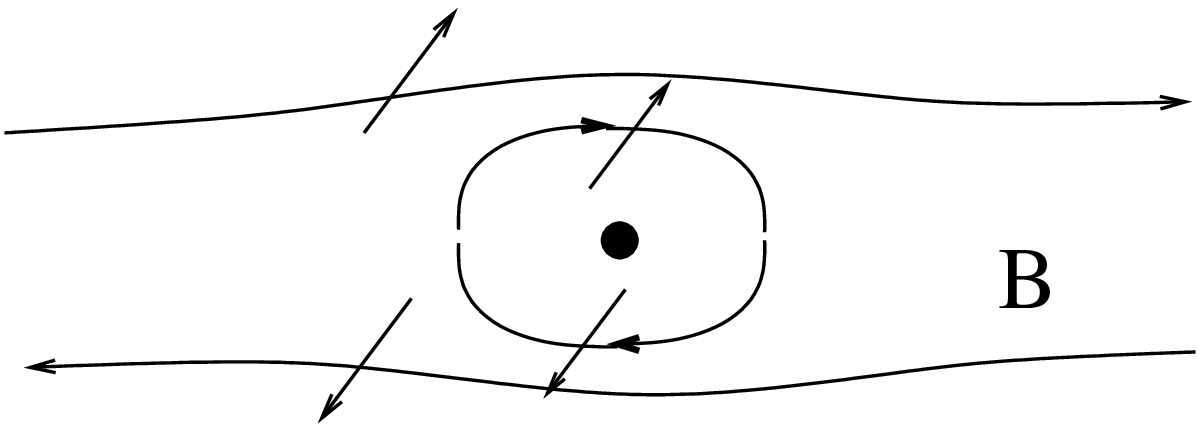,clip=1,width=0.7\linewidth}
\caption{A) The
RG flow of the Kondo problem in the spin Hall edge state. The
antiferromagnetic FP (the dashed line) describes the formation of
the Kondo singlet. The ferromagnetic FP (the solid dot) 
shows a non-zero
critical value of $J_z$. B) Kondo screening spin current vortex
around the local moment.}\label{fig:kondo}
\end{figure}

Next we consider the Kondo coupling 
between the local impurity and edge states, which reads
\bea
H_{K}&=&\int dx \delta (x) \Big \{\frac{J_\parallel}{2}
(\sigma_- \psi^\dagger_{R\uparrow} \psi_{L\downarrow}+ \sigma_+
\psi^\dagger_{L\downarrow}
\psi_{R\uparrow} )\nonumber \\
&+& J_z \sigma_z (\psi^\dagger_{R\uparrow}\psi_{R\uparrow}
-\psi^\dagger_{L\downarrow}\psi_{L\downarrow})\Big \},
\label{eq:kondo}
\eea
where $\sigma_{\pm}=\sigma_x\pm i\sigma_y$,
and $\sigma_z$ denotes the spin 1/2 local moment. $J_\parallel$
describes the spin-flip process accompanied by the single particle
backscattering between the left and right channels, while $J_z$ is
the non-flipping process with forward scattering of electrons.
Combining Eq. \ref{eq:kondo}, the helical liquid
Hamiltonian Eq. \ref{eq:ham0}, and Eq. \ref{eq:fw} together, the
standard RG analysis \cite{anderson1970,Kane1992} gives
\bea
\frac{d J_z}{d \log L/a}= 2 J^2_\parallel,  \ \ \ \frac{d
J_\parallel}{d \log L/a}= (1-K+2 J_z) J_\parallel, \label{eq:rg}
\eea
where $L$ is the infra-red length scale. Eq. \ref{eq:rg} can
be integrated: $J_\parallel^2-(J_z-(K-1)/2)^2=c$ (c: constant).
A nonzero value of $K-1$ contributes to the RG equation at the
tree level due to the anomalous scaling dimension of the
backscattering term. Consequently, as shown in Fig. \ref{fig:kondo} A,
the RG flow pattern is shifted
as a whole by an amount $(K-1)/2$ along the $J_z$ direction
compared to the usual case \cite{anderson1970}.
The strong coupling antiferromagnetic fixed
point (FP) at $J_z\approx J_\parallel\rightarrow +\infty$ still
means the Kondo singlet formation.
It is remarkable that
with the repulsive forward scattering,  i.e., $K<1$, the Kondo
singlet can still form even with a weak ferromagnetic Kondo
coupling. After the formation of the Kondo singlet, it  behaves
like a spinless impurity, which can only cause a phase shift to the
edge electrons.

The Kondo singlet in the spin Hall edge exhibits a new feature
different from that in the non-chiral systems. According to
Nozieres's strong coupling picture \cite{nozieres1980}, an
electron from the conduction band is bound around the local moment
to form a local singlet. However, the edge electron here only has
one helicity.
Thus the spin flip process inside the Kondo
singlet is accompanied with the backscattering of the edge
electrons. As a result, the screening electron is actually
``circling" around the local moment as depicted in Fig.
\ref{fig:kondo}. B. In realistic systems, the width of the edge states
is still finite. Thus the Kondo screening cloud can be considered
as a spin current vortex, and we speculate that its orbital angular
momentum is quantized.

In conclusion, we have shown that the edge states of the recently
proposed quantum spin Hall systems form a helical liquid,
which is a new class different from the spinless or chiral Luttinger liquid.
We have shown that this liquid with an odd number of components
can only arise as the edge of a 2D system
and we have analyzed its stability under interactions,
impurity scattering and
disorder. The new feature of the Kondo singlet was also studied.
While there is no strict $Z_2$ topological distinction between the
helical liquids with an even or odd number of components
in the presence of strong interactions, the latter
is robust in practice against disorder and interactions.

{\bf Note added:} After the completion of our work, we learned
that the problem of two particle scattering in a helical liquid due to
quenched disorder, has also been independently studied by Xu and
Moore \cite{XU2005}.

We thank T. L. Hughes and C. L. Kane for  helpful discussions.
C.W. and B.A.B. acknowledge support from the Stanford Graduate
Fellowship. This work is supported by the NSF under grant numbers
DMR-0342832 and the US Department of Energy, Office of Basic
Energy Sciences under contract DE-AC03-76SF00515.
C. W. is also supported by the NSF under Grant No. PHY99-07949.


\end{document}